\documentclass[acmsmall]{acmart}
\AtBeginDocument{%
  }

\setcopyright{acmlicensed}
\copyrightyear{2025}
\acmYear{2025}
\acmDOI{XXXXXXX.XXXXXXX}

\acmISBN{978-1-4503-XXXX-X/2025/02}

\begin{document}

\title{Assignment 3: Photoshop Batch Rendering Using Actions for Stylistic Video Editing}

\author{Tessa De La Fuente}
\email{alexandria538@tamu.edu}
\orcid{1234-5678-9012}
\affiliation{%
  \institution{Texas A\&M University}
  \city{College Station}
  \state{Texas}
  \country{USA}
}

\renewcommand{\shortauthors}{De La Fuente}

\begin{teaserfigure}
  \centering
  \includegraphics[width=1\textwidth]{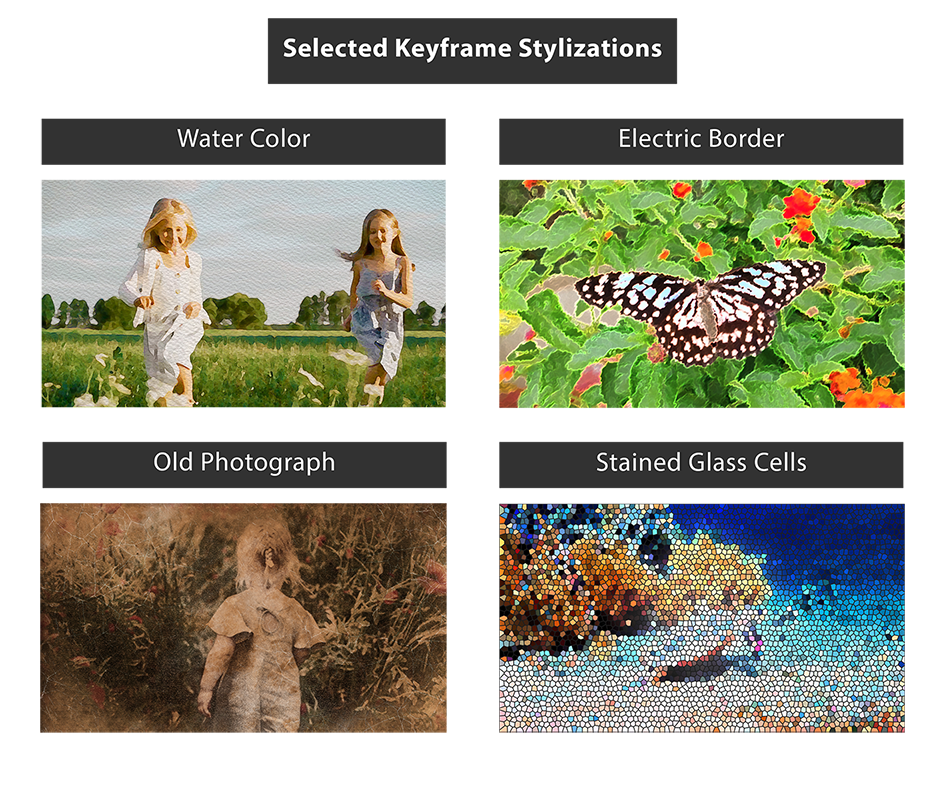}
  \caption{These are four stylistic edits I did through the use of Photoshop’s Batch Render System using Actions.}
  \Description{Four stylistically edited video frames: Water Color, Electric Border, Old Photograph, and Stained Glass Cells}
  \label{fig:teaser}
\end{teaserfigure}

\maketitle


\section{Abstract}
  My project looks at an efficient workflow for creative image/video editing using Adobe Photoshop Actions tool and Batch Processing System. This innovative approach to video editing through Photoshop creates a fundamental shift to creative workflow management through the integration of industry-leading image manipulation with video editing techniques. Through systematic automation of Actions, users can achieve a simple and consistent application of visual edits across a string of images. This approach provides an alternative method to optimize productivity while ensuring uniform results across image collections through a post-processing pipeline.

\section{Introduction and Related Works}
As a digital artist passionate about video/creative editing with movies like Wolf Walkers, The Mitchells vs The Machines and Loving Vincent as inspiration [Figure 2], I've found Adobe Photoshop to be an invaluable tool for both video and image manipulation. While traditionally known as an image editing tool, using “...Batch to apply Photoshop actions to batches of images” offers powerful advantages for video editing workflows (Eismann, pg. 16). An action, defined as “a small file of recorded instructions…[which] can be 'played' on images,” can be saved and easily replicated by an artist, enabling consistent, highly customizable, time-efficient processing of multiple video frames while maintaining style and artistic integrity (Nichols, pg. 26). Photoshop is geared towards “photographers of all skill levels,” and thus, some may find it more comfortable to dabble in creative video editing (Nichols, pg. Xiii). While “there is more advanced video editing software such as Adobe Premiere Pro and Apple’s Final Cut…this software contains more features and is more complex”, and I experimented with this process because I think Photoshop has merit to be considered as effective video editing tool and may be used in conjunction with such programs to help workflows at least in rapid prototyping or if users are unfamiliar with certain programs (David, pg. 23).

\newpage
\begin{figure}[h]
  \centering
  \includegraphics[width=0.9\textwidth]{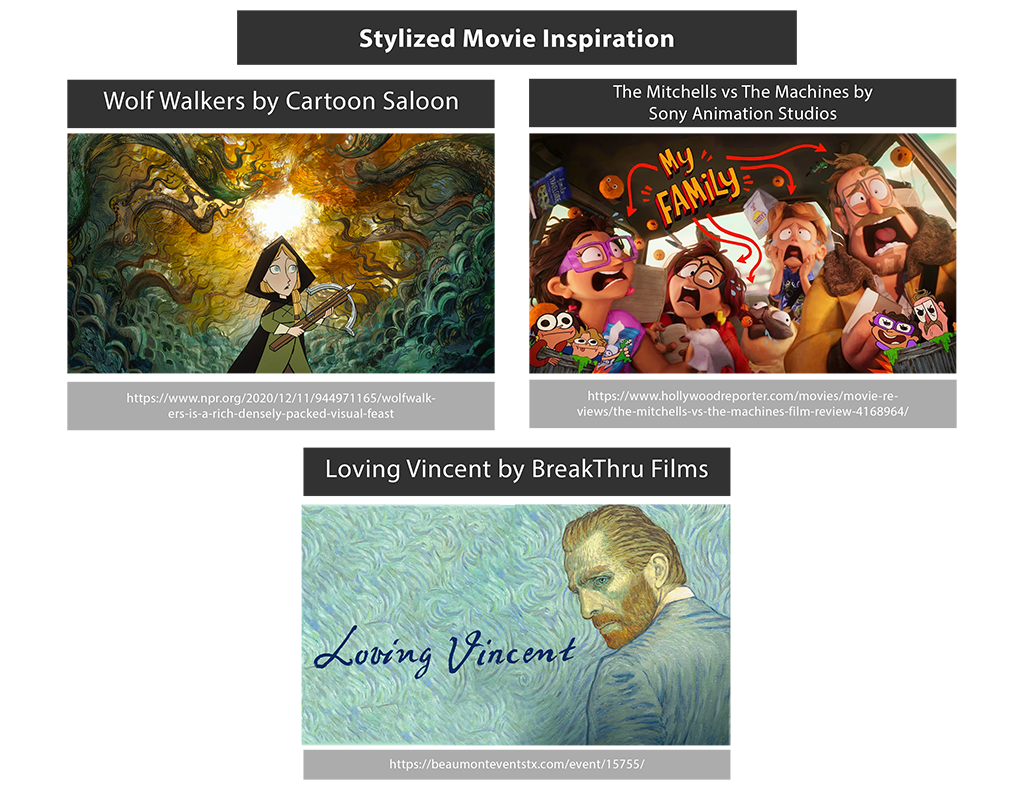}
  \caption{Figure 2. These are three movies with uniquely creative styles and artistic techniques that inspire me in my exploration of stylistic editing and experimental animation.}
  \Description{Three images from stylized movies: Wolf Walkers by Cartoon Saloon, The Mitchells vs The Machines by Sony Animation Studios, and Loving Vincent by BreakThru Films.}
\end{figure}

\section{Methodology}
My method for testing the capabilities of Photoshop for creating video or group image editing relied on two primary Photoshop features: the Actions Tool and the Batch Render Tool [Figure three]. Starting with royalty-free videos (or alternatively shot footage), I selected four slow-motion clips so that the motion wouldn’t be too distracting from the style alterations. I tried to make sure there was a clear focus and showed decent contrast and clarity to be visually interesting. [Figure four]. Each video was imported into Photoshop and transformed into a sufficient number of frames for smooth transitions [Figure five]. After saving the files, I created four creative styles to apply using tools like Filter Gallery, Stylize, Camera Raw filter, and Adjustment Layers. Documenting each step was crucial in order to efficiently replicate the process in an action recording. To do this, open a keyframe, go to the Action Panel>Create New Action in the Actions pane, and repeat the effects you wish to use [Figures 6-9]. Importantly, the save step must be included in the action recording. The Batch tool (File>Automate>Batch) was then used to apply these actions across folders, processing each file in two to ten seconds, depending on complexity [Figure 10]. Finally, I imported the processed sequences into Premiere Pro, handling possible naming sequence reversals by organizing images within bins before adding them to sequences and rendering them out. The resulting renders show distinctive style alterations but still retain the elements and movement from the original videos [Figure 11]. To demonstrate versatility, I applied multiple styles to different videos [Figure 12], showcasing both customization options and action re-usability across various clips.

\begin{figure}[h]
  \centering
  \includegraphics[width=0.8\textwidth]{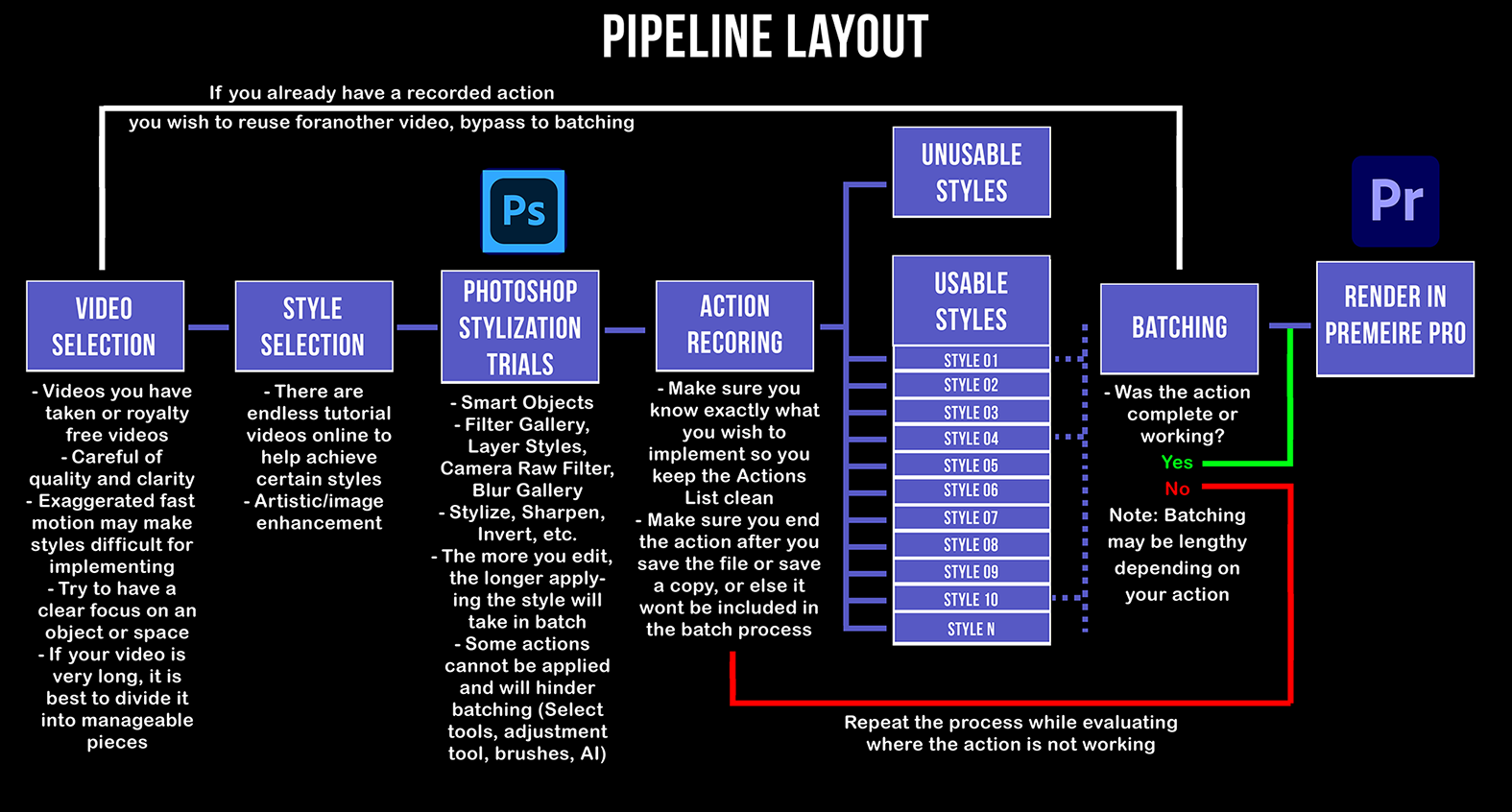}
  \caption{This is the Pipeline I created using Photoshop’s Batch Render System and Actions Tool, along with Premiere Pro for rendering image sequences. Once a video and style are selected, the artist goes through stylization trials to figure out the best method to achieve the desired style. Then the artist can use Action recording to apply to batch rendering a folder of files. If you already have an Action saved and a new video to stylize, you can bypass to the Batching Phase. If the Batch doesn’t run or the resulting image isn’t identical to your vision, return to the Action Recording phase and evaluate what part of the recording isn’t working.}
  \Description{The Pipeline Layout I developed for stylistic editing in Photoshop}
\end{figure}

\begin{figure}[h]
  \centering
  \includegraphics[width=0.85\textwidth]{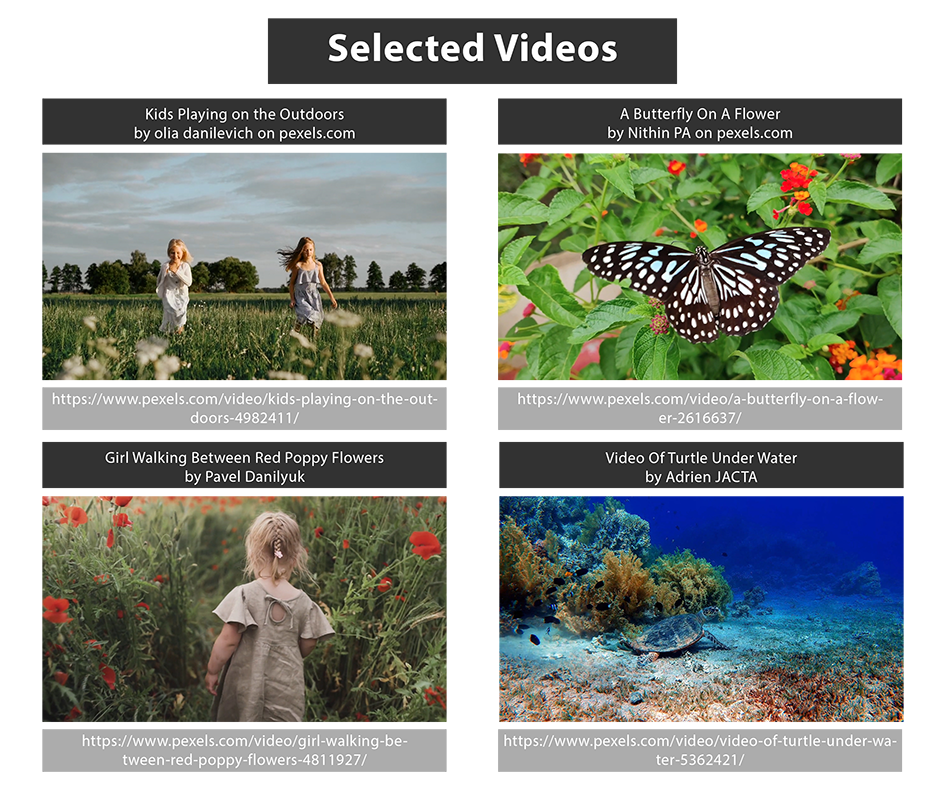}
  \caption{These are the videos I selected from Pexels.com for the trials.}
  \Description{Four images to show the selected videos.}
\end{figure}

\newpage
\begin{figure}[h]
  \centering
  \includegraphics[width=0.8\textwidth]{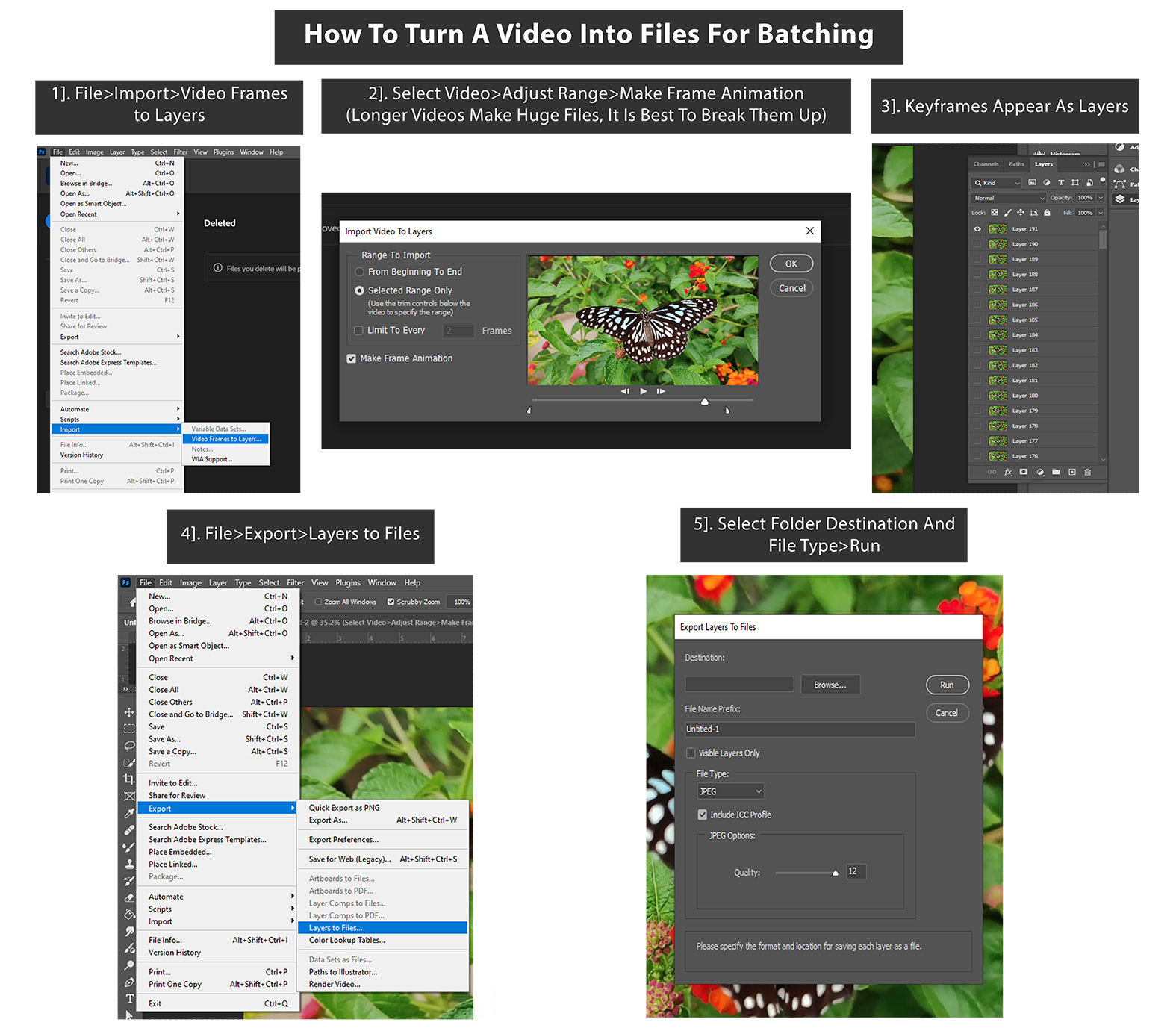}
  \caption{This figure outlines the step-by-step process to import a video into Photoshop and have it export a string of image files for future Batch Rendering.}
  \Description{A figure showing how to turn a video into files for batching in Photoshop.}
\end{figure}

\begin{figure}[h]
  \centering
  \includegraphics[width=0.8\textwidth]{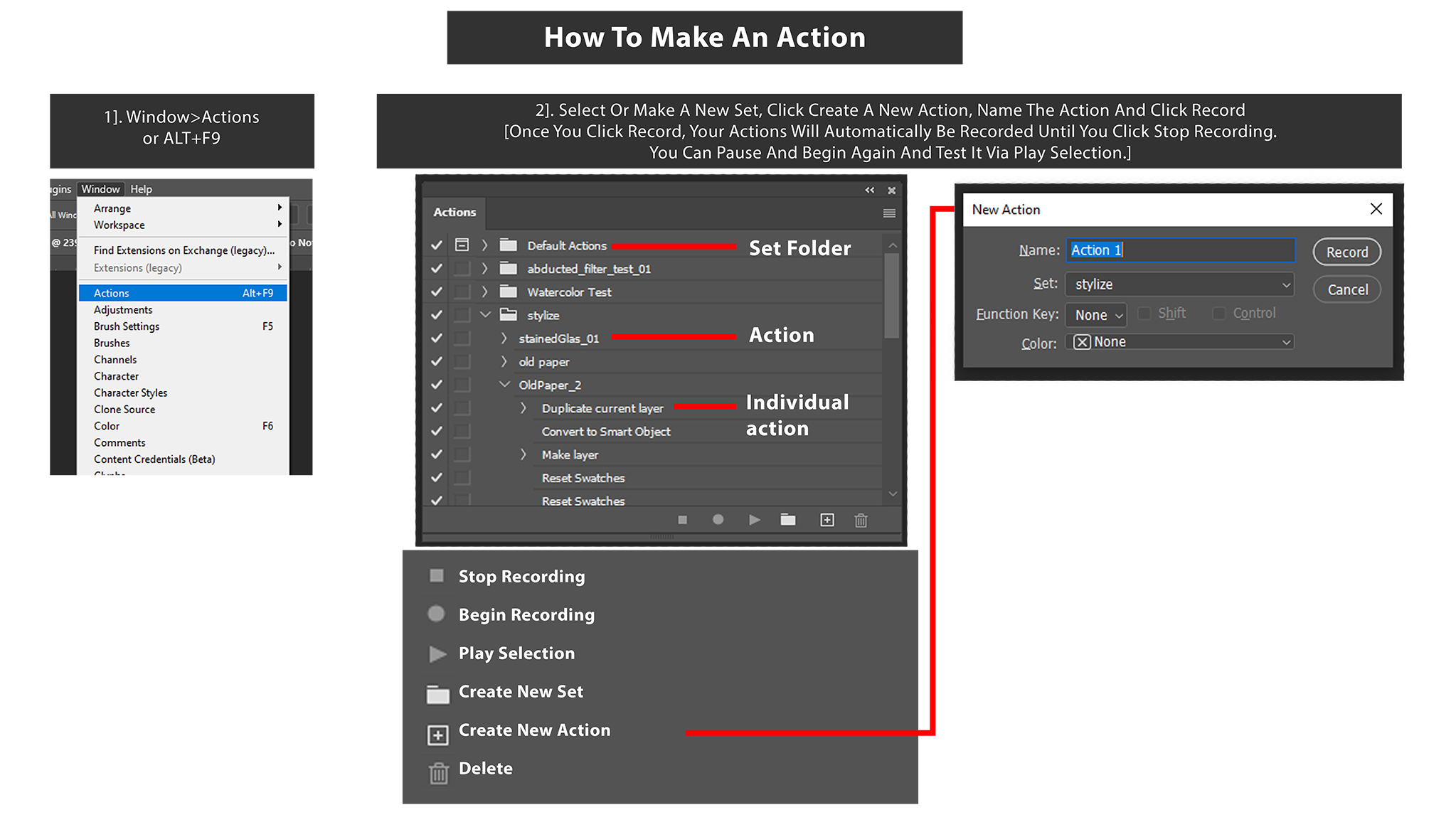}
  \caption{This figure outlines the process for starting an Action to record your image edits.}
  \Description{A figure showing how to make an action in Photoshop.}
\end{figure}

\clearpage

\begin{figure}[h]
  \centering
  \includegraphics[width=0.9\textwidth]{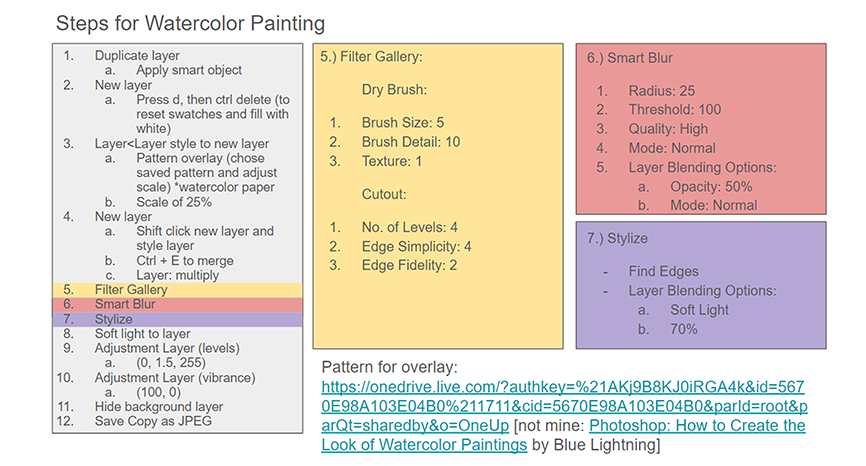}
  \caption{These are the steps to replicate my Action for turning a video keyframe into a watercolor painting. I used this video as inspiration: Geller, Marty. “ Photoshop: How to Create the Look of Watercolor Paintings.” YouTube, Blue Lightning TV, 17 Apr. 2022, www.youtube.com/watch?v=FhgTFAUZrMs. Note that depending on your image, alterations might be necessary for the best effect.}
  \Description{A list of steps for stylistic editing for a Water Color style.}
\end{figure}

\begin{figure}[h]
  \centering
  \includegraphics[width=0.9\textwidth]{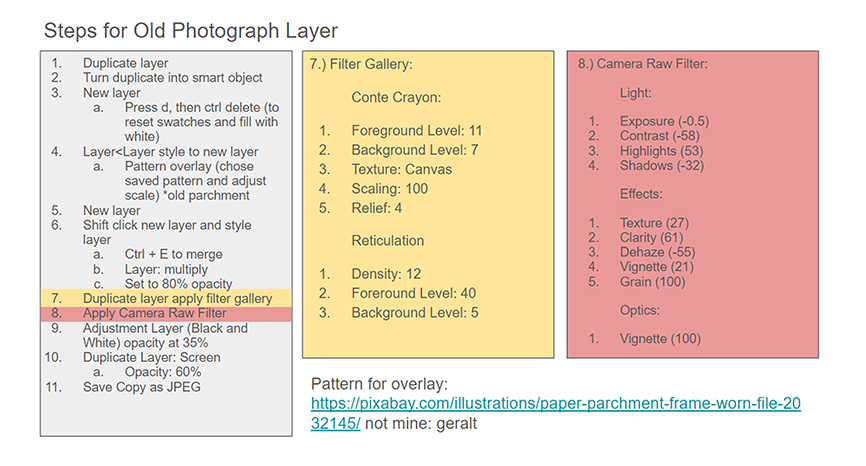}
  \caption{These are the steps to replicate my Action for turning a video keyframe into an Old Photograph. I used this video as inspiration: “Make Any Photo Vintage in Adobe Photoshop | Tutorial.” YouTube, SonduckCreative, 20 Jan. 2023, www.youtube.com/watch?v=Oa5qvjTkCuY. Note that depending on your image, alterations might be necessary for the best effect.}
  \Description{A list of steps for stylistic editing for an Old Photograph style}
\end{figure}

\pagebreak

\begin{figure}[h]
  \centering
  \includegraphics[width=0.8\textwidth]{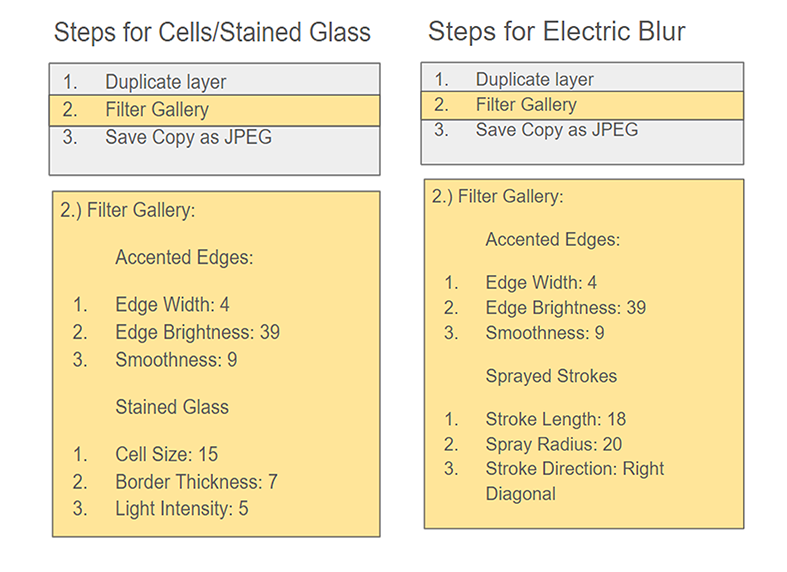}
  \caption{These are the steps to replicate my Actions for stylizing a video keyframe into stained glass and to have an electric border. Note that depending on your image, alterations might be necessary for the best effect. These lean more abstracted for video editing.}
  \Description{A list of steps for stylistic editing for Stained Glass/Electric Blur styles}
\end{figure}

\begin{figure}[h]
  \centering
  \includegraphics[width=0.7\textwidth]{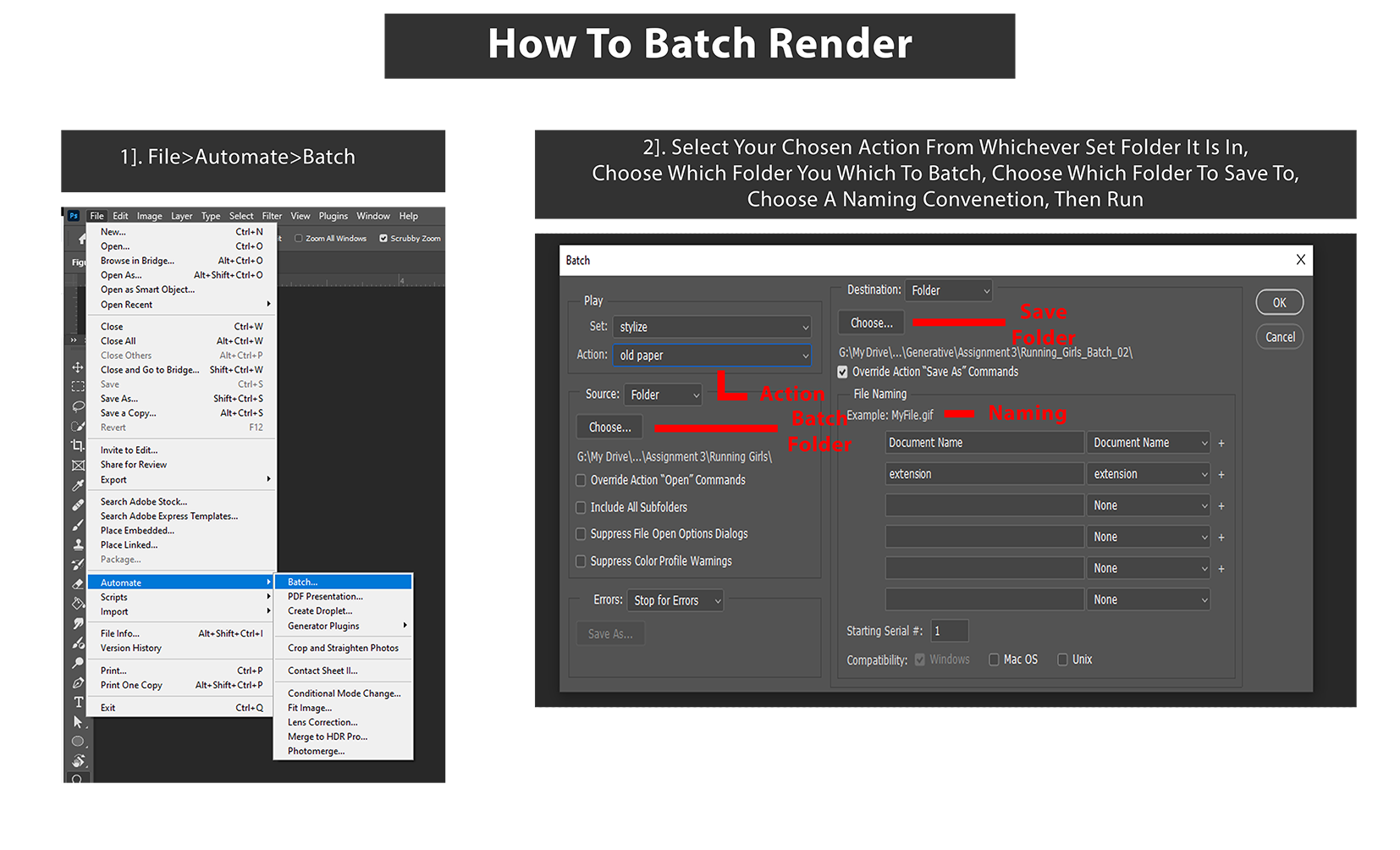}
  \caption{This figure outlines the process for initiating a Batch Render in Photoshop after you have created an Action. I highly recommend enabling “Stop for Errors” in case something is amiss in your recorded Action. I also recommend making your Batch Folder and your Save Folder different.}
  \Description{A figure showing how to batch render in Photoshop.}
\end{figure}

\pagebreak

\section{Result and Future Work}
Here is a Link to the rendered videos:\\ https://drive.google.com/drive/folders/1bbva6FQ2peCfKfe0PDtP9EU0inysZV\textunderscore P?usp=sharing

With the videos I created, I believe I have shown that maintaining consistency and showing a lot of customizability were easily attainable through the use of these Photoshop tools. The process also shows flexibility between file types and easier recovery from errors as the process can be easily replicated or altered based on artistic need. Depending on the effects used, the consistency between frames may be affected by Jittering. I found there are methods to cut down on flickering via tools like deflicker in Premiere Pro (Pueringer). It did arguably take a longer time than video processing programs, but I had already expected that, and I was nonetheless happy with the outcomes of the stylizing. The process of highlighting the usefulness of batch rendering and action recording for video editing, rather than photo grouping alone, I believe, was a success. I would like to explore other tools to add in conjunction with the action tool or other programs that can benefit from the process of quick image stylization. For example, EbSynth, which requires selected keyframes, would benefit from this by selecting the specific files you wish to use and putting them through a relatively small batch render to cut down on time (Secret Weapons).

\pagebreak

\begin{figure}[h]
  \centering
  \includegraphics[width=0.7\textwidth]{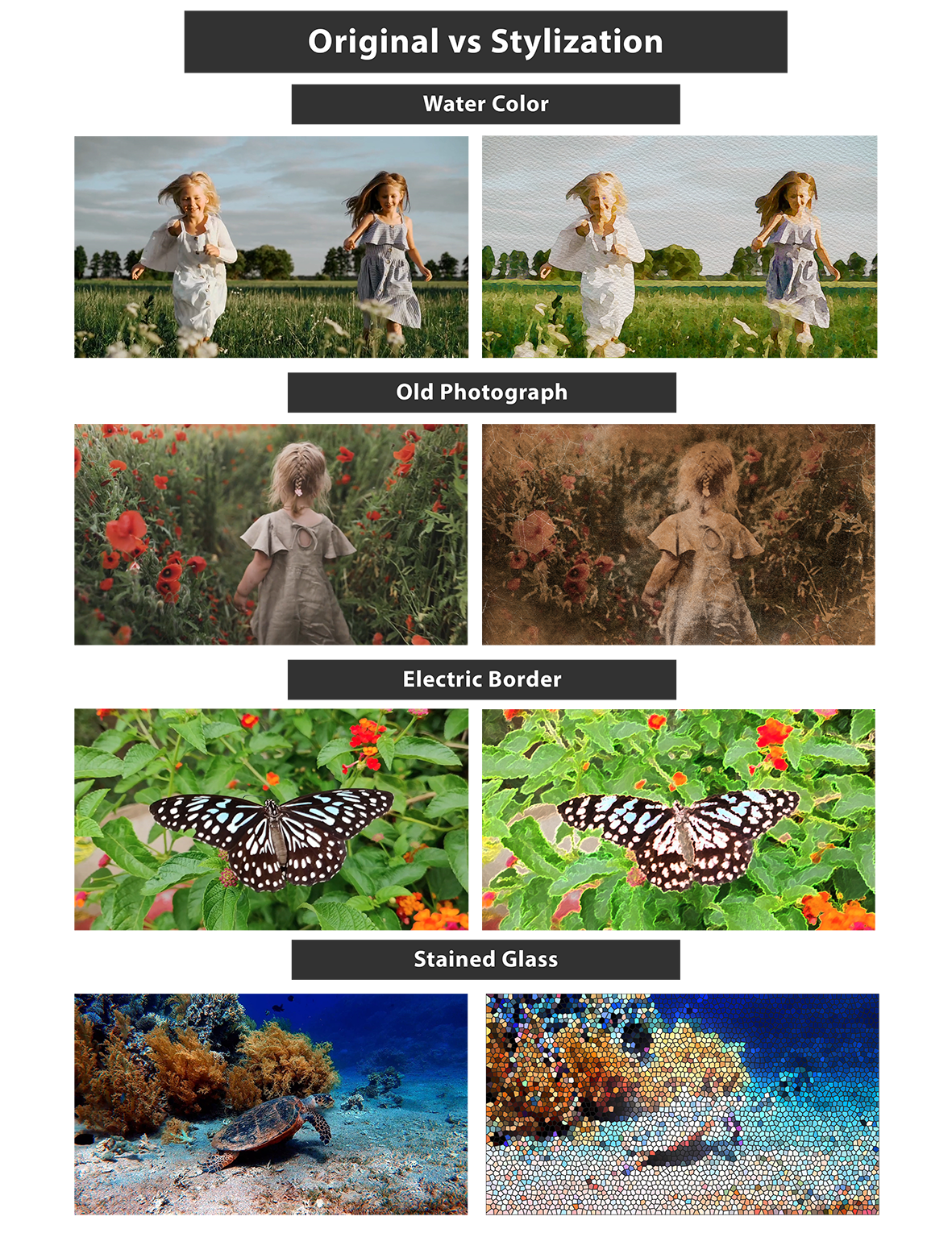}
  \caption{Here are side-by-side comparisons of the original video frames and their stylized edits.}
  \Description{A figure showing original and stylized frames side by side.}
\end{figure}

\pagebreak

\begin{figure}[h]
  \centering
  \includegraphics[width=0.7\textwidth]{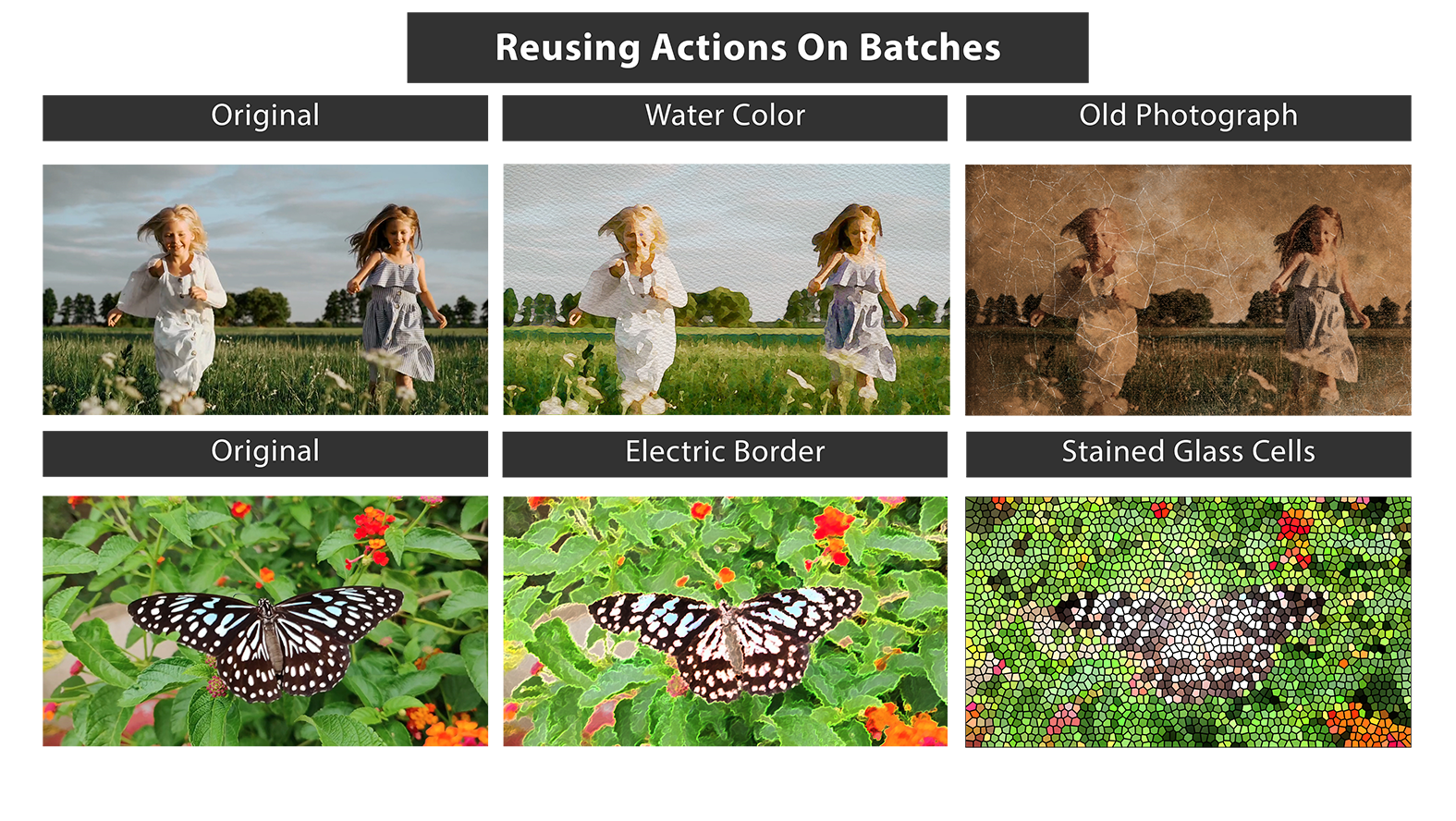}
  \caption{To highlight the re-usability of this process, this figure shows my trials of re-batching image sets with different selected Actions. Note that sometimes Actions need to be altered for the best effect.}
  \Description{A figure showing videos with different styles applied, showing re-usability of this process.}
\end{figure}

\section{Conclusion}
Through this project, I discovered that Photoshop’s Actions and Batch Processing tools offer a uniquely distinctive entry point into image and video editing, valuable for photographers or armature digital artists. While traditional video editors like Premiere Pro excel at timeline-based editing, Photoshop’s frame-by-frame approach provides good control over visual aesthetics. The most memorable aspect was witnessing how you could transform dozens of frames quickly, demonstrating the power of systematic automation in creative workflows. This reinforced that efficiency and creativity aren't mutually exclusive – with the right tools and methodology, artists can maintain high-quality standards while significantly reducing processing time.

\section{Bibliography}

\begin{itemize}
    \item [1] Eismann, Katrin, and Wayne Palmer. Photoshop restoration \& retouching. Peachpit Press, 2006.
    \item [2] Geller, Marty. “ Photoshop: How to Create the Look of Watercolor Paintings.” YouTube, Blue Lightning TV, 17 Apr. 2022, www.youtube.com/watch?v=FhgTFAUZrMs. 
    \item [3] Horowitz, David. "Teaching video editing and motion graphics with Photoshop." Innovative Marketing 13.3 (2017): 17-24.
    \item [4] Nichols, Robin. Mastering Adobe Photoshop Elements 2022: Boost your image-editing skills using the latest Adobe Photoshop Elements tools and techniques. Packt Publishing Ltd, 2021.
    \item [5] Powell, Noel. “Watercolor Painting Effect Tutorial - After Effects.” YouTube, Creation Effects, 11 May 2016, www.youtube.com/watch?v=tix2u3Hg9lA.
    \item [6] Pueringer, Niko, and Dean Hughes. “Did We Just Change Animation Forever?” YouTube, Corridor Crew, 26 Feb. 2023, www.youtube.com/watch?v=\textunderscore9LX9HSQkWo.
    \item [7] Xu, Chengyuan. “Coherent Video Style Transfer Demo (2019, Private).” YouTube, YouTube, 27 Oct. 2022, www.youtube.com/watch?v=nQ\textunderscore itjCBEAs.
    \item [8] “About Actions and the Actions Panel.” Actions and the Actions Panel in Photoshop, 25 Sept. 2023, helpx.adobe.com/photoshop/using/actions-actions-panel.html.
    \item [9] “EbSynth - Tutorial.” YouTube, Secret Weapons, 3 July 2019,\\
    www.youtube.com/watch?v=0RLtHuu5jV4.
    \item [10] “Make Any Photo Vintage in Adobe Photoshop | Tutorial.” YouTube, SonduckCreative, 20 Jan. 2023, www.youtube.com/watch?v=Oa5qvjTkCuY.
    \item [11] “Process a Batch of Files.” Process a Batch of Photoshop Files, 24 May 2023,\\
    helpx.adobe.com/photoshop/using/processing-batch-files.html.
\end{itemize}

\section{Contact}
Email – alexandria538@tamu.edu\\
Personal Website - https://tdelafuente.myportfolio.com/

\begin{acks}
This work is submitted as part of Assignment 3 for the VIZA 626 course at Texas A\&M University, under the instruction of Professor You-Jin Kim, during the Spring 2025 semester.

VIZA 626 Class Website: https://sites.google.com/view/viza626/
\end{acks}


\bibliography{sample-base}

\end{document}